\documentclass[apjl]{emulateapj}

\usepackage{graphicx}
\usepackage{dcolumn}
\usepackage{bm}

\usepackage{wasysym}
\usepackage{amssymb}
\usepackage{yfonts}

\usepackage{lipsum}
\usepackage{xcolor}
\usepackage{multirow}

\begin{document}


\title{Asymmetric Dark Matter imprint on low-mass main-sequence stars in the Milky Way Nuclear Star Cluster}

\author{Jos\'e Lopes}
\author{Il\'idio Lopes}
\affiliation{Centro de Astrof\'isica e Gravita\c{c}\~ao---CENTRA, Departamento de F\'isica,
Instituto Superior T\'ecnico---IST, Universidade de Lisboa---UL, Av. Rovisco Pais 1, 1049-001 Lisboa}

\date{\today}

\begin{abstract}

In this work, we study the impact of asymmetric dark matter (ADM) on low-mass main-sequence stars in the Milky way's nuclear star cluster, where the dark matter (DM) density is expected to be orders of magnitude above what is found near the Sun ($\rho_{\text{DM}} \gtrsim 10^3 \ \text{GeV cm}^{-3}$). Using a modified stellar evolution code and considering a DM particle ($m_{\chi} = 4 \ \text{GeV}$) with a spin-dependent interaction cross section close to the limits allowed by direct detection, we found that the interactions of ADM with baryons in the star's core can have two separate effects on the evolution of these stars: a decrease in the hydrogen burning rate, extending the duration of the main-sequence of stars with $M \sim 1 \ M_{\astrosun}$ by a few Gyr;  the suppression of the onset of convection in the core of stars with $M \lesssim 1.5 M_{\astrosun}$, and consequent quench of supply for the nuclear reactions. If we consider $\rho_{\text{DM}} > 10^3 \ \text{GeV cm}^{-3}$ (corresponding to the inner $~5$ pc of the Milky Way), stars lighter than the Sun will have a main-sequence life span comparable to the current age of the universe. Stars heavier than two solar masses are not sensitive to the DM particles considered here.
\end{abstract}


\maketitle

\section{Introduction}
\label{sec:intro}

Since the beginning of the last century, a remarkable amount of evidence hinting at a dominant contribution of particle dark matter (DM) to the total mass of the universe has been found \citep[][and references therein]{1978ApJ...225L.107R,2010ARA&A..48..495F}. Since then, different detection mechanisms have been devised to probe the elusive nature of DM. Amongst these, the search for manifestations of DM in stars arose as an ingenious and inexpensive indirect detection strategy, taking advantage of the rich diversity of physics found within stars, allied with the extensive understanding we currently possess on the standard picture of stellar evolution. Naturally, in this context, most indirect searches have been focused in the Sun, yielding fruitful results that are competitive and complementary to direct detection experiments \citep{2002MNRAS.337.1179L,Lopes:2010fx,2015PhRvL.114h1302V,2016ApJ...827..130L,2018PhRvD..98l3012A}. A remarkable example is the absence of a neutrino signal from DM annihilation in the Sun, which has placed the strongest limits to date for the spin-dependent interactions of low-mass DM \citep{2013PhRvL.110m1302A,2015PhRvL.114n1301C}. 

As a DM detector, the Sun is a laboratory with unique conditions that cannot be replicated on any man-made experiment. However, its sensitivity is limited by the local DM density, $\rho_{\text{DM}} \simeq 0.39\ \text{GeV cm}^{-3}$ \citep[e.g.][]{2010JCAP...08..004C}, which is orders of magnitude below the density expected near the center of the Milky Way \citep{2017PDU....15...53H}. Luckily or not, this region of the galaxy is also home to a very dense cluster of stars called a nuclear star cluster  \citep[NSC; for a review of NSCs see][]{2010RvMP...82.3121G}, which is composed by a rich diversity of stars, some of them potential probes to the local population of DM particles. However, since the NSC lies on the galactic disk plane, observations in the optical waveband are obscured by the interstellar dust permeating the line of sight to the galactic center. Moreover, the luminosity contamination by bright massive stars populating the NSC poses a challenge to the observation of fainter less massive stars \citep{1990MNRAS.244..706A,1991ApJ...382L..19K}. Nonetheless, recent developments in adaptive optics and infrared (IR) imaging and spectroscopy in telescopes such as ESO's Very Large Telescope have dramatically improved the angular resolution of observations to approximately $50 \ \text{mas}$ (corresponding to a linear resolution of $\sim 2\times10^{-3}$ pc at the center of galaxy). The unprecedented status of observations in the region of the galactic center has not only allowed us to resolve NSC stellar populations with a near-infrared magnitude of $K_s \approx 18$, corresponding to several Gyr old 1--2$M_{\astrosun}$ stars, but also to study of the diffuse light of unresolved populations with $K_s$ = 19--22, most likely composed by 0.8--1.5 $ M_{\astrosun}$ main-sequence (MS) stars \citep{GallegoCano:2017ht,2018A&A...609A..27S}. In the next 5--10 years, so-called extremely large telescopes with approximately $40$ m diameter mirrors should be able to resolve these fainter stars and allow the construction of color-magnitude diagrams and estimation of the age and metallicity of the stellar populations not only in the Milky Way, but also in the NSC of galaxies as far as $\sim 5 \ \text{Mpc}$ \citep{2014A&A...568A..89G}. The promising outlook for future detailed observations of stars in NSCs emphasizes the importance of understanding the effects of DM--nucleon interactions in the center of the star and the impact that these effects can have on the evolution of MS stars embedded in environments with very high densities of DM particles, as should be the case in the center of a galaxy. 

In this work we consider a specific class of weakly interacting massive particle (WIMP) models, which is the so-called asymmetric dark matter \citep[ADM; e.g.,][]{2014PhR...537...91Z}, a model motivated by the remarkable similarity between the \textit{a priori} unrelatable DM and baryon relic abundances inferred from cosmic microwave background measurements \citep{2016A&A...594A...1P}. Unlike the general picture of self-annihilating WIMPs produced during the universe freeze-out \citep[e.g.,][]{1986PhRvD..33.1585S}, ADM arises due to a particle-antiparticle asymmetry in a mechanism analogous to baryogenesis. This parallel between the DM and baryonic sectors would naturally explain the observed relation between the DM and baryonic relic densities, $\Omega_{\text{DM}} \simeq 4.5 \Omega_{\text{B}}$, assuming a DM particle with $m_{\chi} \sim 4.5$ GeV \citep{2009PhRvD..79k5016K}. Models with a negligible self-annihilation cross section like that of ADM are also specially interesting in searches of DM manifestations in stars: the number of particles captured inside the star is not limited by DM self-annihilation, increasing the star's sensitivity to a possible DM signature. Furthermore, unlike DM models with non-negligible self-annihilation, the content of ADM in the NSC is not constrained by null results from indirect searches that look for byproducts of DM annihilation in the center of the galaxy \citep{Gondolo:1999bt,2014PhRvL.113o1302F,2017PhRvD..96f3008L}.

In the next section we describe the framework used throughout this work, after which we present the results on the effects of interactions between ADM and baryons in the stellar medium in section \ref{sec:evac}. Section \ref{sec:evol} is dedicated to the discussion of the impact that these effects can have on the properties and evolution of the stars, and how can we use them to look for a convincing ADM signature. In section \ref{sec:conc} we conclude with important final remarks. 

\section{Energy transport by ADM}

Due to their massive nature, ADM particles permeating the Milky Way can get captured inside stars, which can lead to the buildup of large concentrations of particles within the center of the star. Differing from standard WIMPs with a non-negligible annihilation cross section, ADM particles will not self-annihilate, and through interactions with matter in the stellar plasma, will provide an extra way to carry energy from hotter to colder regions of the star. If the number of ADM particles captured in the star is sufficiently high, the extra transport of energy can have a non-negligible impact on the structure of the star \citep{2010Sci...330..462L,2019PhRvD..99b3008L,Taoso:2010df}.

Assuming that self-interactions between ADM particles are negligible, the number of particles inside the star at any given time $t$ is defined by
\begin{equation} 
\frac{\text{d}N_{\chi}}{\text{d}t} (t) \simeq C_{\chi}(t),
\label{eq:adm_number}
\end{equation}
where $C_{\chi}$, the ADM capture rate, is a function of $t$. In fact, eq. \ref{eq:adm_number} is only valid when evaporation, the inverse process of capture, is negligible, which, in the scenarios considered here, is true for particles with $m_{\chi} \gtrsim 3 \ \text{GeV}$ \citep{1990ApJ...356..302G,1987ApJ...321..560G}. The capture rate $C_{\chi}(t)$, first described by \cite{1987ApJ...321..571G}, is mainly dependent on the particle's mass $m_{\chi}$, interaction cross section with each nucleon $\sigma_{\chi,n_i}$, DM density $\rho_{\text{DM}}$ and the chemical composition of the star. As such, $C_{\chi}(t)$ is approximately constant during the MS \citep{Lopes:2011ce}. 

In this work we consider ADM particles with $m_\chi = 4 \ \text{GeV}$, which not only is consistent with the observed relic densities $\Omega_{\text{DM}}$ and $\Omega_{\text{B}}$ (see sec. \ref{sec:intro}), but is also the mass for which the ADM interactions in the solar plasma are maximal. This can by explained by the fact that $m_\chi = 4 \ \text{GeV}$ is the closest value to the mass of hydrogen (the main chemical element present inside MS stars) for which it is safe to assume that evaporation is negligible. 

On the other hand, the nature of the interactions with nucleons is highly model-dependent. In most direct and indirect searches, it is typical to consider that non-relativistic DM elastic scatterings with nucleons can be described by an effective constant scattering cross section. Furthermore, it is standard procedure to consider spin-dependent and spin-independent contributions separately. In this work we are interested in the case where scatterings are predominantly spin-dependent, $\sigma_{\text{SD}} \gg \sigma_{\text{SI}}$ and we fix the spin-dependent cross section to $\sigma_{\text{SD}} = 10^{-37} \ \text{cm}^{-2}$, which is approximately the maximum cross section not in disagreement with other detections experiments for $m_{\chi} \simeq 4 \ \text{GeV}$ \citep[e.g.,][]{2017PhRvL.118y1301A,2017APh....90...85B,2015PhRvL.114n1301C}.

For such an ADM particle\footnote{In the scope of this work, the presence of a non-negligible self-annihilation cross section is the only difference between ADM and the more usual class of thermally produced WIMPs. Despite the fact that the energy transport formalism described here was developed considering self-annihilating WIMPS, it is also valid for the specific case of ADM, and for that reason, in this work we eeuse the term ``ADM particles'' rather than ``WIMPs'' to refer to the DM particles.}, the mean-free-path in the center of the star, $\ell_{\chi,0}$ in a typical MS star is several times larger than the typical scale radius of the ADM distribution in the star core, which is given by
\begin{equation}
r_{\chi} = \left( \frac{9}{4 \pi} \frac{k_{\text{B}}T_c}{G \rho_c m_\chi}\right)^{\frac{1}{2}}.
\label{eq:wimp_rad}
\end{equation} 
This means that after scattering with a nucleon, an ADM particle will orbit the star several times before interacting with the stellar plasma again. In this regime, the so-called Knudsen limit, the Knudsen number, defined by $K \equiv \ell_{\chi,0} / r_\chi$, is much larger than one. Differently from the case where $K \ll 1$ \citep{Gilliland:1986jx,1990ApJ...352..654G}, energy transport by ADM in the Knudsen limit is non-local, and thus its treatment is highly non-trivial. \cite{1985ApJ...294..663S} obtained an analytical approximation for the ADM energy transport in this limit, assuming two interacting Maxwellian gases of different temperatures. However, using a Monte Carlo method to solve the Boltzmann collision equation, \cite{1987ApJ...321..560G} later showed that the isotropy assumption implicit in the Maxwellian approximation overestimates the ADM energy transport capability by a factor of a few. Furthermore, \cite{1987ApJ...321..560G} also showed that the Monte Carlo results for the Knudsen limit could be somewhat replicated assuming the ADM luminosity in the conduction limit (i.e. when $K \ll 1$) multiplied by a suppressing factor. 

To study these effects during the evolution of MS stars, as well as the extent at which each treatment produces different results, we extended the publicly available Modules for Experiments in Stellar Astrophysics (MESA) stellar evolution code \citep{2011ApJS..192....3P,2013ApJS..208....4P,2015ApJS..220...15P,2018ApJS..234...34P} to include ADM capture and energy transport in a robust and consistent way. In both formalisms described in the last paragraph, energy transport by ADM particles can be treated as an additional term in the energy balance of the star,
\begin{equation}
\varepsilon_{\chi} = \frac{\partial}{\partial m} l_{\chi},
\label{eq:transport}
\end{equation}
where $m$ is the position in the star in the Lagrangian description and $l_{\chi}$ is the ADM luminosity. Because ADM transports energy within the star, the energy term in eq. \ref{eq:transport} is negative in regions where the heat is carried away from (typically in regions where there is energy being produced in nuclear reactions), and positive where it is deposited. Also, assuming that evaporation of ADM from the star is negligible (as it is the case here), the summation of $\varepsilon_{\chi}$ throughout the star should yield zero given that there is no net outflow of energy from the star.

We implemented both energy transport treatments, i.e., the analytic approximation by \cite{1985ApJ...294..663S} and the suppressed conductive limit proposed by \cite{1990ApJ...352..669G}. Studying different scenarios of ADM impact on MS stars, we found that while the former slightly overestimates $\varepsilon_{\chi}$ in comparison with the latter, only the former is numerically stable for scenarios when the ADM energy contribution in a given region is comparable to the energy produced in nuclear reactions, i.e., when $\left|\varepsilon_{\chi}\right| \simeq \left|\varepsilon_{\text{nuc}}\right|$. For this reason, and to be able to obtain predictions for the cases when $\varepsilon_{\chi} \simeq \varepsilon_{\text{nuc}}$, we decided to use the the analytic approximation by \cite{1985ApJ...294..663S}. It should be noted that, while this treatment might indeed overestimate the energy evacuation for a given ADM scenario, the actual functional form and signal of $\varepsilon_{\chi}$ as a function of the star mass profile is similar for both approaches, and thus the phenomenology resulting from the effects studied here is independent of the implementation strategy.

\section{Energy evacuation by Dark Matter}
\label{sec:evac}
Following the formalism described in the last section, we studied the effects of ADM energy transport within MS stars in the Milky Way's NSC. To study the effects of ADM in MS stars of the NSC we evolve all models from the pre-main-sequence to the beginning of the red giant Branch considering an ADM density of $\rho_{\chi} = 10^3 \ \text{GeV cm}^{-3}$. Assuming that the ADM density profile in the Milky Way follows a Navarro--Frenk--White profile \citep[e.g.,][]{Fornasa:2014kj}, the ADM density considered here corresponds to a galactocentric distance of $r_{\text{GC}} \simeq 5$ pc, which is approximately the effective half-light radius of the Milky Way's NSC \citep{2011ASPC..439..222S}. It should be noted that because the ADM density increases as we approach the center of the Milky Way, the value for $\rho_{\chi}$ considered here is in fact a conservative estimation. In some cases we also present results for $\rho_{\chi} = 10^2\ \text{GeV cm}^{-3}$, which corresponds to $r_{\text{GC}} \simeq 50$ pc.

Given the rich diversity of physics found across the mass range of stars in the MS, it is expected that ADM should have different effects on the evolution of stars with different mass. While stars with mass closer to the Sun ($M \lesssim 1.2 \ M_{\astrosun}$) are fueled by stable hydrogen burning proton-proton (pp) chain reactions, stars with $M \gtrsim 1.2 \ M_{\astrosun}$ can develop a convective core during the MS due to the higher output of energy produced in the dominant carbon--nitrogen--oxygen (CNO) cycle. Furthermore, both the pp chain reactions and the CNO cycle are highly temperature-dependent and thus the homogenization of the central temperature profile promoted by ADM energy transport can change the rate of energy production, and consequently, the pace of the star's evolution in the MS. These effects can be concurrent, so it is important to study them separately within the relevant star mass ranges. Stars heavier than $2 \ M_{\astrosun}$ will not be sensitive to the model considered in this work, as the ADM energy transport is negligible when compared to the energy produced in nuclear reactions. All models are evolved with an initial helium abundance of $Y = 0.28$ and have solar-like metallicity ($Z = 0.020$). Convection is treated according to the mixing length theory \citep{1958ZA.....46..108B} as formulated by \cite{1968pss..book.....C} with $\alpha_{\text{MLT}} = 2.0$.

\begin{figure}[!htb]
	\centerline{
		\includegraphics[width=1.0\columnwidth]{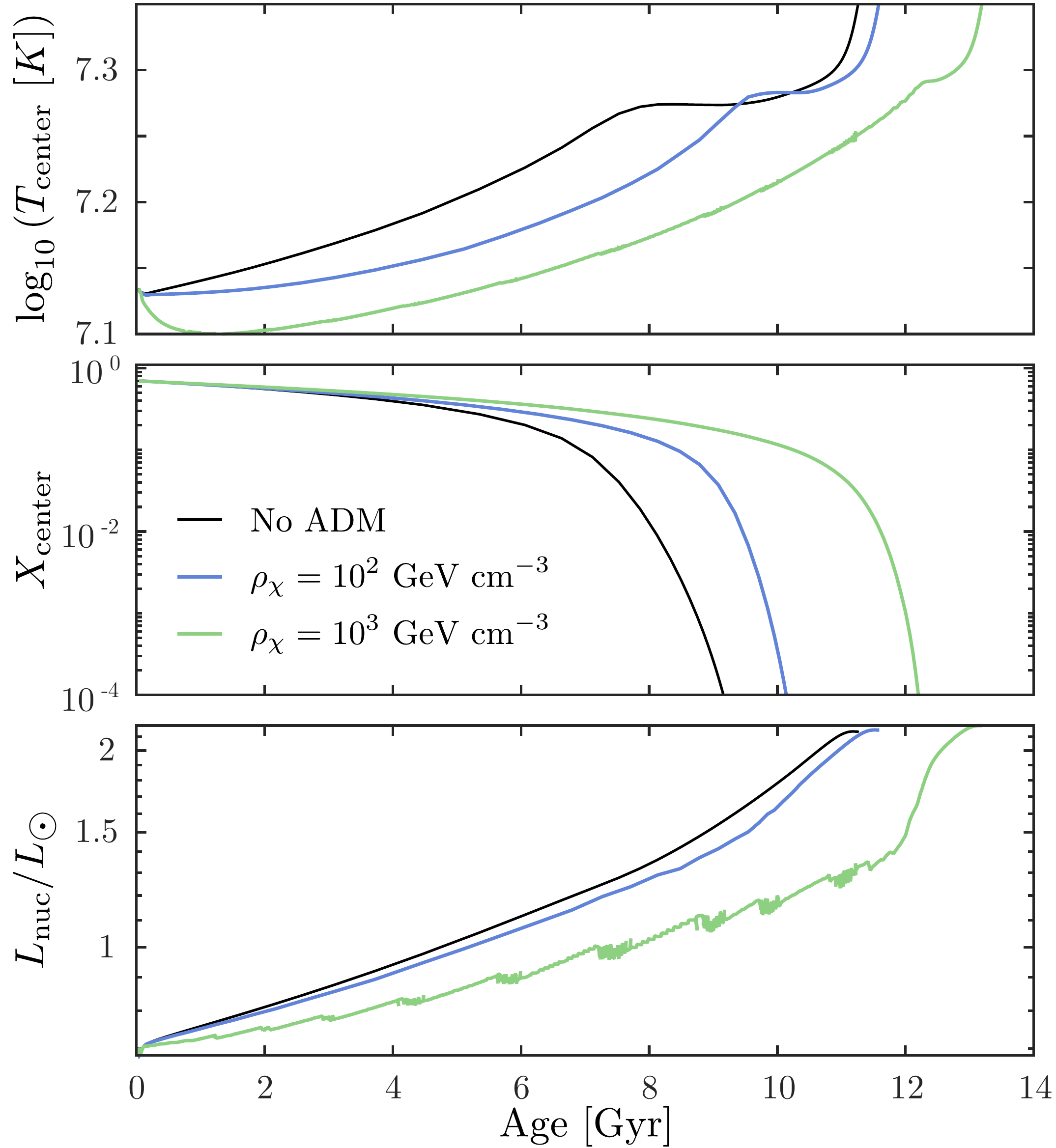}}
	\caption{Central temperature, central hydrogen abundance and total luminosity from nuclear reactions, respectively, of a $1 M_{\astrosun}$ MS star in different ADM density scenarios ($m_{\chi} = 4 \ \text{GeV}$).}
	\label{fig:coolness}
\end{figure}

\subsection{Main-sequence turnoff}
\label{sec:turnoff}

The time a star spends in the MS is mainly defined by the supply of hydrogen in the core and the rate at which it is burned. In stars with $M < 1.2 \ M_{\astrosun}$ hydrogen is predominantly burned through the pp chain reaction for most of the MS life span. As described earlier, captured ADM particles will increase the plasma's capacity to transfer heat from hotter to colder regions, distributing the energy within the typical scale radius of the ADM distribution in the star (see eq. \ref{eq:wimp_rad}). The distribution of energy promoted by ADM interactions will flatten the center of the stellar temperature profile, cooling the region where the nuclear reactions are taking part. Since pp reactions are moderately temperature-dependent, this will slow down the rate of nuclear reactions, which will take longer to exhaust the hydrogen supply in the core, resulting in a longer MS phase. This can be seen in fig. \ref{fig:coolness} where the evolution of a star with $M = 1.0 \ M_{\astrosun}$ is shown in different ADM scenarios. Assuming the moment when the central content of hydrogen reaches $X_c = 10^{-4}$ as representative of the end of the MS (and $t_{\text{MS}}$ as the time the star spent until reaching that point), while the standard case with no ADM has $t_{\text{MS}}\simeq9$ Gyr, the same star, evolved in an environment with $\rho_{\chi} \simeq 10^2 \ \text{GeV cm}^{-3}$, will have $t_{\text{MS}}\simeq10$ Gyr. The case when $\rho_{\chi} \simeq 10^3 \ \text{GeV cm}^{-3}$ is even more dramatic, with $t_{\text{MS}}\simeq12 $ Gyr, which is comparable to current estimations of the universe age, $\simeq 13,5$ Gyr. 

\begin{figure*}
	\centerline{
		\includegraphics[width=2.0\columnwidth]{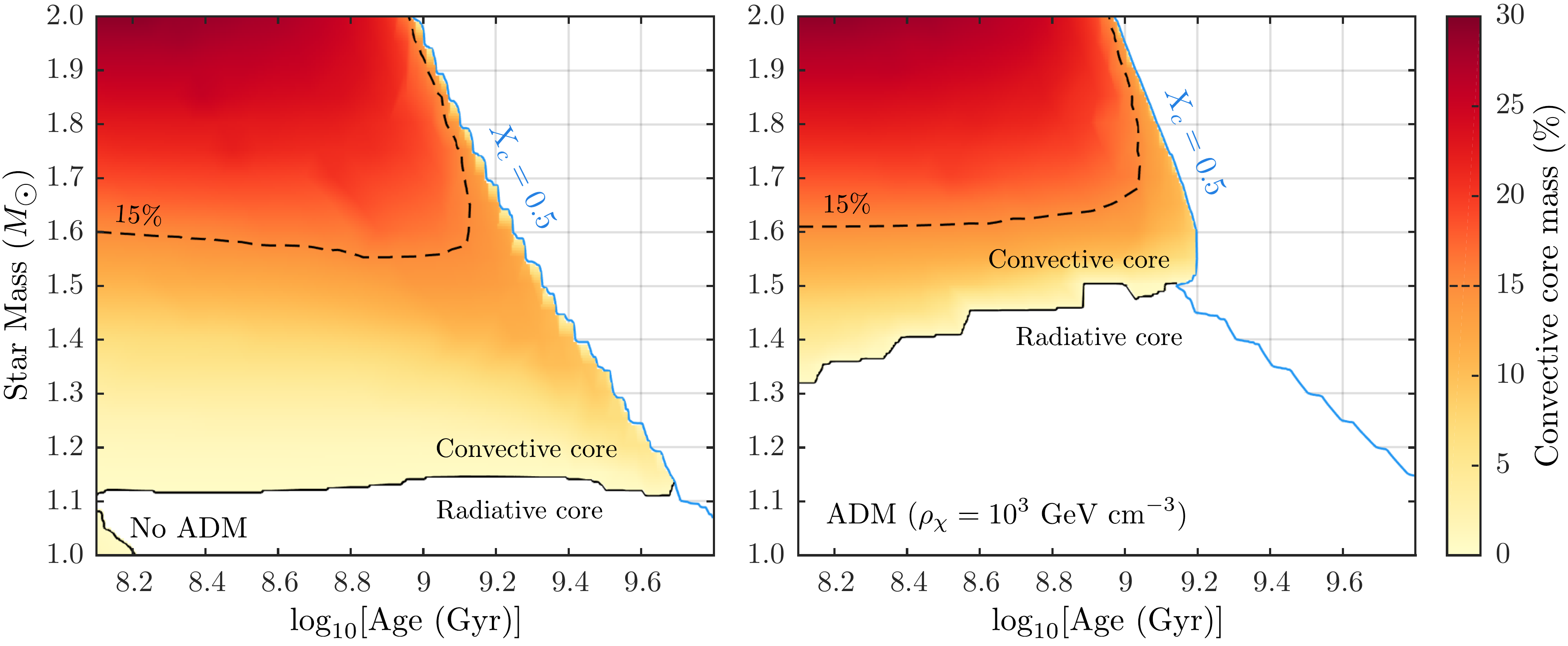}}
	\caption{The mass of the convective core during the MS of stars in the mass range $1M_{\astrosun}<M<2M_{\astrosun}$ for scenarios with $\rho_{\chi} \simeq 10^3 \ \text{GeV cm}^{-3}$ (right) and without ADM (left). The blue line is representative of the end of the MS, which here we define as the stage when hydrogen in the center of the star is at $X_c = 10^{-4}$. The black solid line separates stars with radiative cores from stars with convective cores. We also show the contour (black dashed line) for which the mass of the convective core represents 15\% of the total mass of the star.}
	\label{fig:conv_supp_2d}
\end{figure*}

This result is particularly interesting because it could explain the current deficit in the observed number of red giant and horizontal branch stars within the inner parsec of the NSC \citep{Do:2009kz,2016ApJ...823..155K}. Rather than being a shortcoming of the luminosity contamination by the large density of bright young massive stars, the low number of late-type low-mass stars in the Milky Way NSC observations could be explained by the extension of the core hydrogen burning phase caused by ADM energy transport illustrated in fig. \ref{fig:coolness}.

\subsection{Convective core suppression}
\label{sec:conv_supp}

The delay of the MS phase shown in fig. \ref{fig:coolness} will be less pronounced for stars with higher mass, mainly due to the increase in nuclear energy output that comes with the increase in the star's central temperature. For stars heavier that the Sun ($M \gtrsim 1.2 \ M_{\astrosun}$, depending on the chemical composition of the star) the CNO cycle will take the place of the pp chain reaction as the main contributor to the total energy produced in nuclear reactions. The CNO cycle will not only be more sensitive to the star's central temperature but also produce more energy than the pp reaction chain, with $\varepsilon_{\text{CNO}} \sim T^{18}$ compared to $\varepsilon_{\text{pp}} \sim T^4 $ \citep{2011RvMP...83..195A}. If the temperature gradient in the core is too steep and the stellar plasma is too opaque to the radiative transfer of the large output of energy created in the CNO cycle, the core will ``boil'' and convection will take over as the predominant energy transport mechanism in this region, evacuating energy through the displacement of portions of hot stellar plasma to colder regions of the star. The instability of a given region of the star against convection, known as the Schwarzschild criterion, is defined by
\begin{equation}
\nabla_{\text{rad}} = \frac{3}{16 \pi a c G} \frac{P}{T^4} \frac{\kappa l}{m} > \nabla_{\text{ad}},
\end{equation}
where $\kappa$ is the plasma opacity, $\nabla_{\text{ad}} \equiv \text{d log }T/\text{d log }P$ is the adiabatic gradient and $l$ is the luminosity, which measures the energy flux passing through the relevant region. The onset of convection in the center of the star will define its structure, because convection will promote the mixing of chemical elements within the convective core.

\begin{figure}[!htb]
	\centerline{
		\includegraphics[width=1.0\columnwidth]{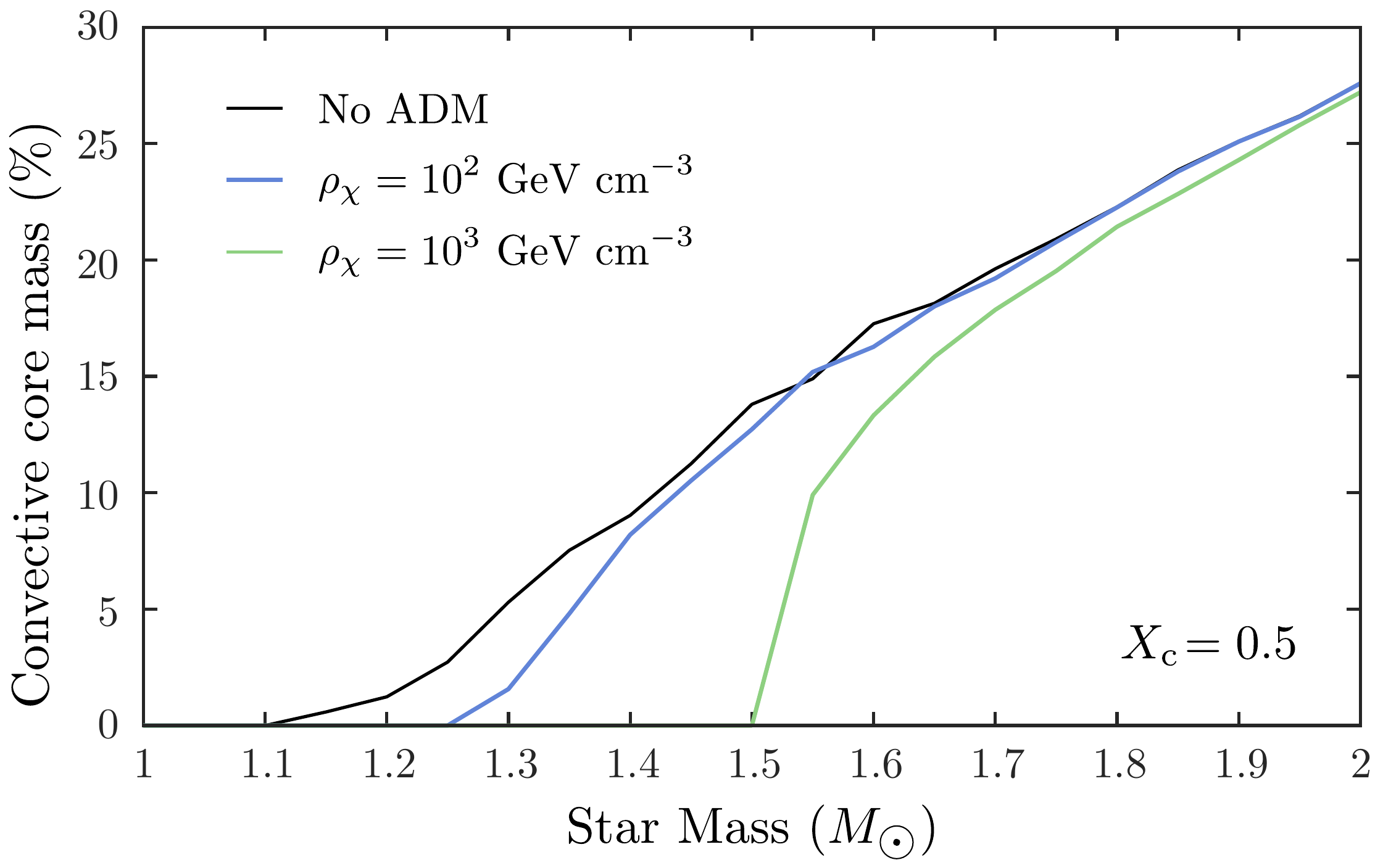}}
	\caption{Mass of the convective core when the central hydrogen abundance is 0.5. The conditions within the star during the moment when $X_{\text{c}} = 0.5$, represented here, are representative of the conditions during the MS lifetime (see fig. \ref{fig:conv_supp_2d})}
	\label{fig:conv_supp}
\end{figure}

Considering ADM capture and energy transport, the ADM energy term (see eq. \ref{eq:transport}) will be negative within the hotter regions of the stellar plasma and will balance the energy produced in the nuclear reactions, contributing to the stability of the region against convection. If the absolute value of $\varepsilon_\chi$ is comparable with the energy produced in nuclear reactions, energy transport by ADM particles will suppress the onset of convection. This is shown in fig. \ref{fig:conv_supp_2d}, where the size of the convective core of a star during the MS as a function of the total mass of the star is shown in the classical scenario (i.e., with no ADM) and considering ADM capture and energy transport with $\rho_{\chi} = 10^3 \ \text{GeV cm}^{-3}$. As shown, in the case with no ADM, stars with $M \gtrsim 1.1 M_{\astrosun}$ will have a convective core during the entire duration of the MS. On the other hand, if we consider energy evacuation by ADM, MS stars with $M \lesssim 1.4 M_{\astrosun}$ will have a radiative core during the largest part of their MS lifetime. This result is also in agreement with results obtained by \cite{2015PhRvD..91j3535C}. The difference between the different scenarios decreases as we consider heavier stars, because the effects of $\varepsilon_\chi$ in the energy balance between convection and radiation are mitigated by the increase in the energy produced in nuclear reactions. This result is emphasized in fig. \ref{fig:conv_supp}, where we compare the mass of the convective core of different stars when they have reached a central hydrogen abundance of $X_c = 0.5$, a phase representative of the rest of the MS. Defining $M^{\text{conv}}_{0}$ as the minimum star mass for which convection arises in the core, we can see that $M^{\text{conv}}_{0}$ increases as we consider higher ADM density. On the other hand, the size of the convective core in stars with $M > 2.0 M_{\astrosun}$ is not sensitive to ADM energy transport due to the large energy flux produced in the CNO cycle. Nonetheless, for the ADM density expected within the Milky Way NSC ($\rho_{\chi} \simeq 10^3 \ \text{GeV cm}^{-3}$) the core will be stable against convection for stars as massive as $M \simeq 1.5 \ M_{\astrosun}$. Hence, the observation of a 1.5 $M_{\astrosun}$ MS star in the NSC with a radiative core would be a strong hint to the presence of a large population of ADM particles in the center of the galaxy.

\section{Impact on the evolution}
\label{sec:evol}

\begin{figure}[!htb]
	\centerline{
		\includegraphics[width=1.0\columnwidth]{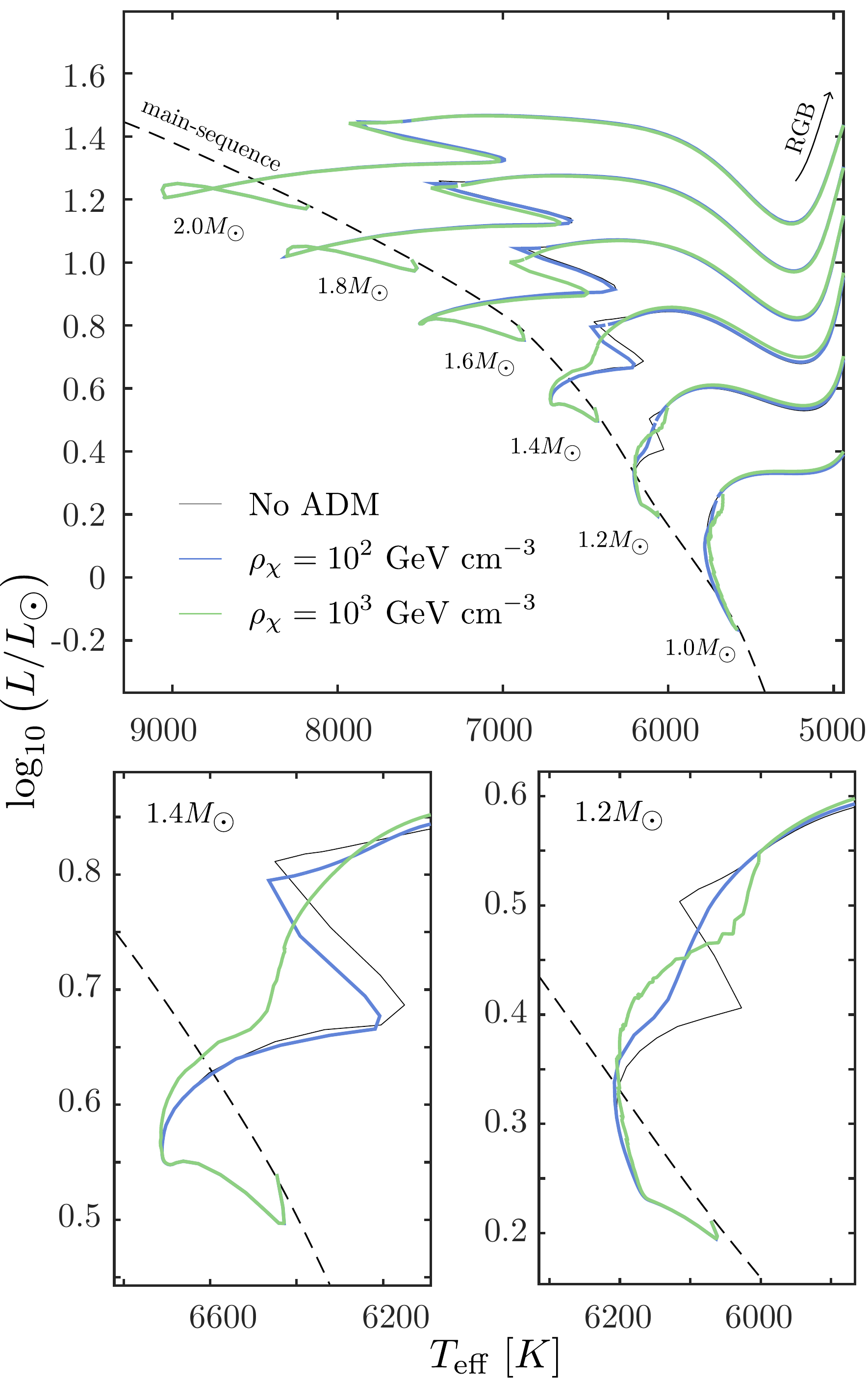}}
	\caption{\textbf{Top}: HR diagram with MS stars evolved in an ADM halo with different ADM densities. All stars start from the zero age main-sequence, and travel all the way up until the red giant branch.\textbf{Bottom}: HR diagram of $1.4M_{\astrosun}$ (left) and $1.2M_{\astrosun}$ (right) stars. All models have solar metallicity.}
	\label{fig:hr_all}
\end{figure}

Both effects related to ADM energy transport explored in sections \ref{sec:turnoff} and \ref{sec:conv_supp} (i.e., the extension of the MS and the suppression of core convection) can have an impact on the evolution of the NSC population of low-mass stars during the MS. Whereas the former prevents stars lighter than the Sun from reaching the red giant phase on a cosmic time-scale, the latter should leave an imprint on the evolutionary tracks of NSC stars with $M \lesssim 2 \ M_{\astrosun}$. In fact, if we look at the evolutionary path of MS stars, one of the most striking features in the HR diagram is the hook in the effective temperature seen for stars with $M \gtrsim 1.2 \ M_{\astrosun}$ (see fig. \ref{fig:hr_all}). This feature is typical of stars that develop a convective core during the hydrogen core burning phase: at the end of the MS, the hydrogen is exhausted within the whole convective core due to the chemical mixing and homogenization caused by convection, leading to an abrupt stop of energy production within the star's core. To mitigate the drop in central temperature caused by the complete halt in nuclear reactions, the star will undergo an overall contraction. The temporary decrease in the radius will increase the star's effective temperature, until the point when hydrogen burning reignites in the shell surrounding the inert helium core, resetting the star on the evolutionary path leading to the red giant branch. In the case of MS stars with non-convective cores, the transition from hydrogen core to shell burning occurs smoothly and thus the hook feature caused by the sudden increase in effective temperature is absent from their evolutionary path.

This can be seen in fig. \ref{fig:hr_all}, where we show the HR diagram of low-mass MS stars in different ADM scenarios. As shown, the evolutionary path of stars with $M \lesssim 1.5 \ M_{\astrosun}$ evolved within an environment with $\rho_{\chi} = 10^3 \ \text{GeV cm}^{-3}$, will not exhibit the hook feature characteristic of core convection, as opposed to the scenario with no ADM where the same feature is visible for stars with $M \simeq 1.2 \ M_{\astrosun}$. Analogously, for $\rho_{\chi} = 10^2 \ \text{GeV cm}^{-3}$ the hook feature is suppressed in the evolutionary tracks of stars with $M \lesssim 1.3 \ M_{\astrosun}$, as highlighted in the lower panels of fig. \ref{fig:hr_all}.

\begin{figure}[!htb]
	\centerline{
		\includegraphics[width=1.0\columnwidth]{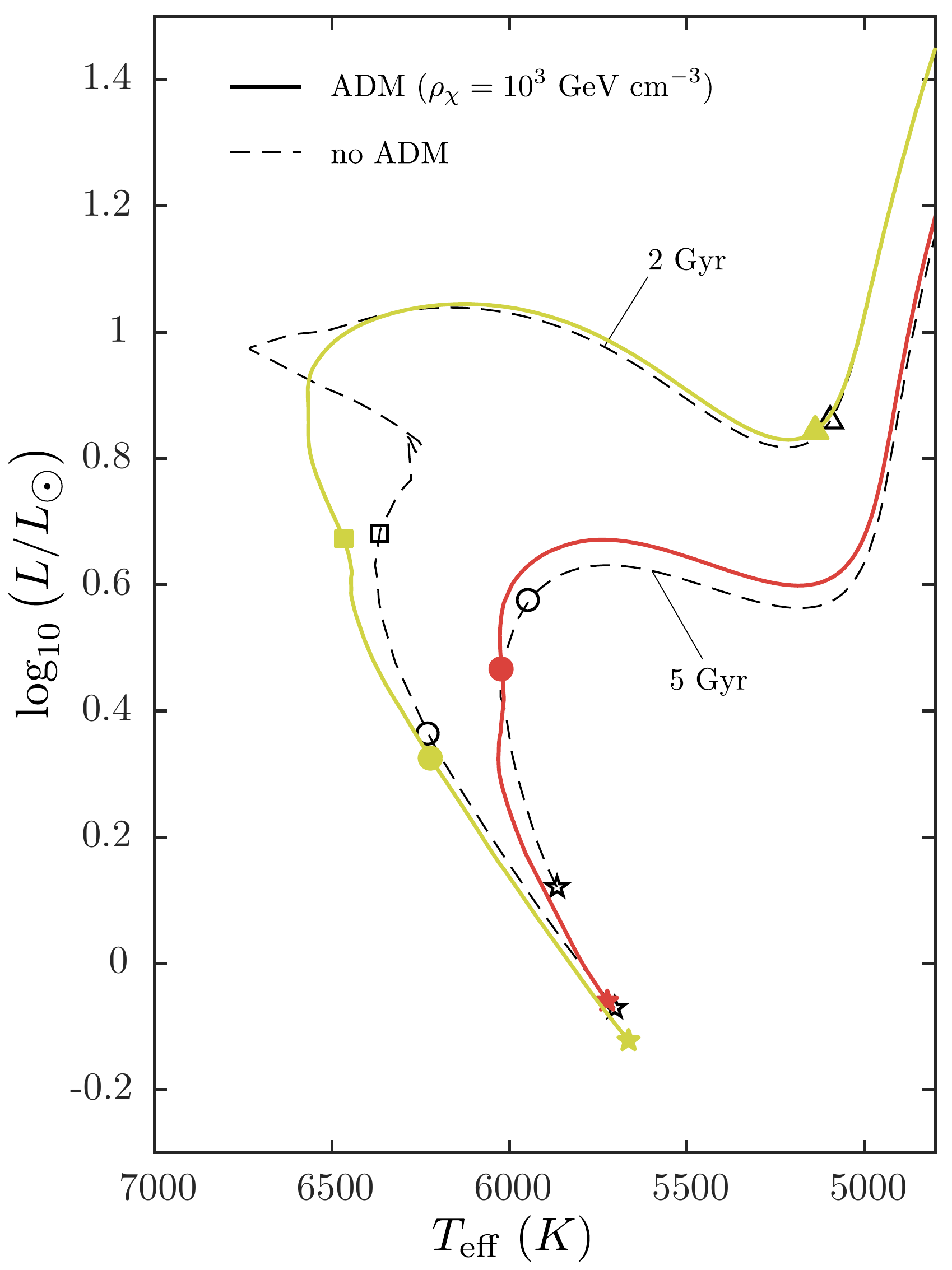}}
	\caption{Isochrones with ages of 2 Gyr and 5 Gyr computed for a stellar population in the mass range $M=1.0-2.0 M_{\astrosun}$, in the classical scenario (no ADM, dashed) and in the case where the cluster evolved in a halo of ADM with $\rho_{\text{DM}} = 10^3 \ \text{GeV cm}^{-3}$ (solid). The isochrones were computed using MIST \citep{2016ApJ...823..102C,2016ApJS..222....8D}. The markers along the isochrones, $\bigstar$ (star), $\bigcirc$ (circle), $\Box$ (square) and $\bigtriangleup$ (triangle), represent stars with 1.0, 1.2, 1.4, and 1.6 $M_{\astrosun}$ respectively. All models have solar metallicity.}
	\label{fig:isochrones}
\end{figure}

The suppression of the MS hook in the HR diagram shown in fig. \ref{fig:hr_all} can be important in the study and analysis of the color-magnitude diagram of a cluster's stellar population. In fact, today, the best estimates of the age of open stellar clusters are obtained by fitting isochrones to the respective color-magnitude diagram, a method in which the identification of the end of the MS, the so-called MS turnoff, plays a key role \citep[e.g.][]{2010ARA&A..48..581S}. If the stars in the turnoff region of a stellar cluster with a given age have convective cores, the respective color-magnitude diagram will display a gap in luminosity corresponding to the hook in the HR diagram. For this reason, isochrone fitting can also be used to probe the convective core of MS stars, as done by \cite{2004PASP..116..997V}, who used this technique to constrain convective overshooting for stars with $M \approx 1.3 M_{\astrosun}$, which, as shown in fig. \ref{fig:hr_all}, is in the mass range for which ADM can dictate whether the core is stable or unstable against convection \citep{2004PASP..116..997V}. 

It should be noted that while isochrone fitting might be a valid approach in the estimation of the age and metallicity of open stellar clusters, the same may not be strictly true for NSCs. While the stellar populations comprising the former can be assumed to be a simple stellar population, i.e., a population composed of stars with roughly the same age and same metallicity, little is known about mechanisms of stellar formation in NSC's. Today it is commonly accepted that the NSC is not only composed by stars created during repeated episodes of in situ star formation but also by stars from clusters that have formed elsewhere and were attracted to the center of the galaxy \citep{2018A&A...609A..28B}. Nonetheless, even though both scenarios might contribute to the uncertainty of parameter estimation by isochrone fitting, a stellar population composed by multiple clusters or with an expanded time formation history can be considered as a superposition of simple stellar populations.

The potential of isochrone fitting as a diagnostic of the ADM content in the NSC is illustrated in fig. \ref{fig:isochrones}, where we show the isochrones for 2 and 5 Gyr. While the overall effect of ADM energy transport is evident in both isochrones, the convective core suppression is especially evident in the 2 Gyr isochrone, where the luminosity gap due to the MS hook is missing for $\rho_{\text{DM}} = 10^3 \ \text{GeV cm}^{-3}$. This striking difference is not present in either much younger or much older isochrones. Clusters younger than 1 Gyr will have a turnoff defined by stars heavier than $2 M_{\astrosun}$, which are not sensitive to ADM energy transport. On the other hand, the turnoff in clusters older than $\sim 4$ Gyr is composed of stars with radiative cores, regardless of the considered ADM density. Nonetheless, in the case of older clusters, energy transport by ADM will introduce a systematic error in the estimation of the cluster age due to the extension of the star's MS lifetime as discussed in section \ref{sec:turnoff}. This delay of the MS turnoff is already noticeable in the 5 Gyr isochrone of fig. \ref{fig:isochrones} (notice the markers along the curves, which represent stars with the same mass). However, while ADM energy transport in stars with $\sim 1 M_{\astrosun}$ extends the core hydrogen burning phase (approximately 3 Gyr for a one solar mass), the same is not true for stars where convection would develop if no ADM was present. The onset of convection in the center of a star will mix the chemical species within the core, providing a continuous supply of burning material to the nuclear reactions during the MS. The suppression of convection by ADM will quench the hydrogen supply, resulting in the early end of the MS phase. 

\begin{figure}[!htb]
	\centerline{
		\includegraphics[width=1.0\columnwidth]{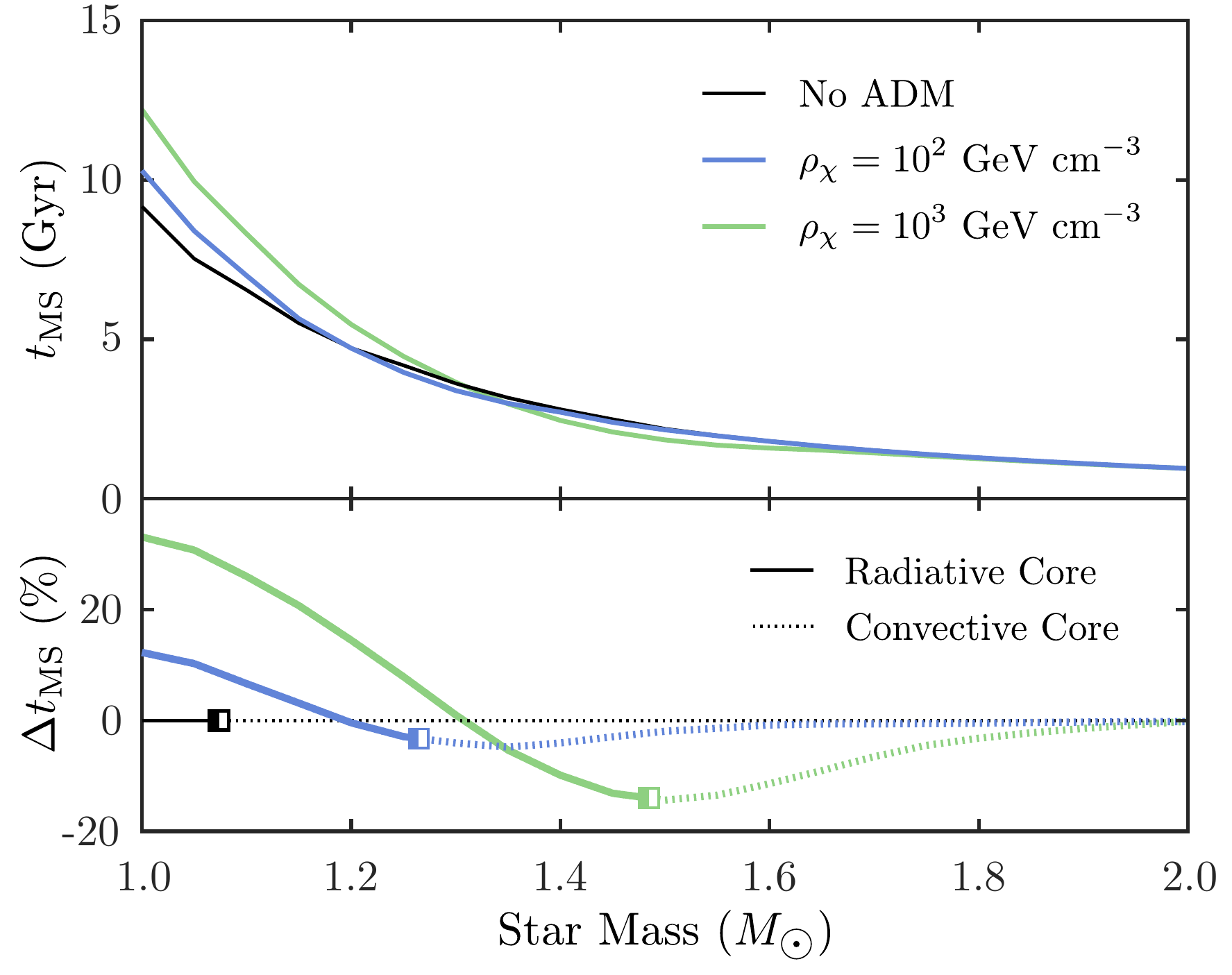}}
	\caption{The time a star spends in the core hydrogen burning stage, which end is here defined as when the hydrogen abundance in the center drops below $X_c = 10^{-4}$, for varying star mass. The bottom panel shows the relative deviation of $t_{\text{MS}}$ with respect to the case with no ADM. The square markers represent the minimum star mass for which the stellar plasma in the core is unstable against convection during the MS.}
	\label{fig:turnoff}
\end{figure}

The combined effects that both the hydrogen burning rate slowdown and suppression of core convection can have on the MS duration of stars in the relevant mass range are shown in fig. \ref{fig:turnoff}. While ADM energy transport with $\rho_{\text{DM}} = 10^3 \ \text{GeV cm}^{-3}$ increases the MS duration of stars with $M < 1.3 M_{\astrosun}$, stars with $M > 1.3 M_{\astrosun}$ will have a smaller $t_{\text{MS}}$ than that in the scenario with no ADM due to convective core suppression. A similar, albeit less pronounced, behavior can be seen for lower ADM density $\rho_{\text{DM}} = 10^2 \ \text{GeV cm}^{-3}$. As expected, stars with $M>2M_{\astrosun}$ will have the same MS duration regardless of the ADM density considered here.

\section{Conclusions}
\label{sec:conc}

In this work, we studied the impact of DM-baryon interactions in low-mass MS stars in the Milky Way's NSC, and how can we explore these effects to probe the elusive nature and content of DM in the center of the Milky Way. We considered ADM interacting with baryons through a spin-dependent effective cross section $\sigma_{\text{SD}} = 10^{-37} \ \text{cm}^{2}$ which sits close to the limits allowed by other experiments for $m_{\chi} \simeq 4 \ \text{GeV}$. 

Assuming a conservative estimation of the DM density expected in the region populated by the NSC ($\rho_{\text{DM}} = 10^3 \ \text{GeV cm}^{-3}$ at $r_{\text{GC}} \simeq 5$ pc assuming a standard Navarro-Frenk-White profile), we found that the transport of energy by ADM inside stars can have two distinct effects: the slowdown of the hydrogen burning rate due to the decrease of the star's central temperature ($\sim$10--20$\%$ lower central temperature in a $1 M_{\astrosun}$ star during the MS) and the suppression of core convection in stars near the limit for which the medium becomes unstable against convection (the minimum star mass for the onset of convection is raised from $M^{\text{conv}}_{0}\simeq1.1M_{\astrosun}$, in the classical case with no ADM, to $M^{\text{conv}}_{0}\simeq1.4M_{\astrosun}$ if we consider ADM). 

Both these effects translate into a non-negligible impact in the evolution of stars with $M<2M_{\astrosun}$. The suppression of the convective core in stars with $M<1.4M_{\astrosun}$ will suppress the MS hook feature in the corresponding HR diagram. From an observational point of view, the absence of the hook feature would result in the suppression of the luminosity gap in the color-magnitude diagram of a $\sim 2$ Gyr old cluster composed by a simple stellar population (i.e., a stellar population with roughly the same age and metallicity), a feature currently observed in multiple open stellar clusters \cite[e.g.,][]{2019MNRAS.484.4718H}. The suppression of the convective core will also quench the hydrogen supply available for nuclear reactions, prompting the stars to leave the MS earlier than they would otherwise (a maximum of $\sim 15 \%$ earlier for $M \simeq 1.5 M_{\astrosun}$). 

It should be noted that in this work we have only considered the simplest stability criterion for convection, the Schwarzschild criterion, with no additional mixing effects such as semiconvection or convective overshooting. While semiconvection is important to account for effects of the molecular weight gradient in the convection criterion \citep{2010Ap&SS.328..129S}, convective overshooting can have an important impact on the extent of the mixing region and thus on the evolution of the star through the MS. In fact, convective overshooting, which has been shown to fit the color-magnitude diagrams of open clusters \citep[e.g.][]{2004ApJ...612..168P}, will increase the mixing region within the stellar core, creating the opposite effect of what is obtained if we consider energy transport by ADM (see fig. \ref{fig:hr_all} and \ref{fig:isochrones}). In this sense, the suppression of the convective core by ADM can be an even stronger diagnostic of the presence of ADM in stars if we consider the effects of overshooting.

On the other hand, stars that are expected to have a radiative core will have a longer MS lifetime ($\sim$ 3 Gyr longer for a $1M_{\astrosun}$ star) due to the decrease in the central temperature, which will slow down the rate of the hydrogen burning rate. Considering this effect, stars lighter than the Sun within environments with $\rho_{\text{DM}} > 10^3 \ \text{GeV cm}^{-3}$ would have a MS lifetime comparable to the current age of the universe, which could help explaining the deficit in the number of observed red giant stars within the inner $\sim$ 0.5 pc of the galaxy \citep{2016ApJ...823..155K,GallegoCano:2017ht}.

In the context of DM indirect searches, the Milky Way NSC is indeed one of the most interesting regions to search for manifestations of particle DM in Stars. The prospect of using these clusters to probe the elusive nature of DM is now more optimistic than ever. Extremely large telescopes, such as the European-ELT and the Thirty Meter Telescope, expected to receive first light within the next decade, will be able to resolve the MS turnoff region of clusters with up to 10 Gyr, not only in the Milky Way's NSC, but also in galaxies as far as 2 Mpc from us. These next-generation ground-based telescopes will blaze a new trail for the exploration of NSC's, including galaxies where the density of DM is expected to be larger than what is found in the Milky Way \citep[e.g.,][]{2012A&A...546A...4T}. Future detailed observations will also shed light on the star formation history and composition of NGCs which, along with a thorough analysis of the effects of DM explored here, should help us understanding the role that DM can have in NSCs, if any.

\begin{acknowledgments}
We thank Bill Paxton for making the MESA code publicly available, as well as Aaron Dotter for the valuable assistance in the construction of the stellar isochrones with the MIST package. We would also like to thank the anonymous referee for the useful comments that have contributed to the discussion of our results. J.L. acknowledges financial support by {Funda\c{c}\~ao para a Ci\^encia e Tecnologia}---FCT Grant No. PD/BD/128235/2016 in the framework of the Doctoral Programme IDPASC---Portugal.
\end{acknowledgments}

\bibliographystyle{aasjournal} 
\bibliography{imprint}
\end{document}